\documentclass[aps, prl, reprint, twocolumn, superscriptaddress, amsmath, amssymb,nofootinbib,
  nobibnotes]{revtex4-2}
\usepackage{graphicx}
\usepackage{amssymb}
\usepackage{color}
\usepackage{booktabs}
\usepackage{subfig}
\usepackage{comment}
\usepackage{appendix}
\usepackage{braket}
\usepackage[colorlinks=true, allcolors=blue]{hyperref}
\usepackage[capitalize, nameinlink]{cleveref}
\usepackage{xr-hyper}
\usepackage{xcolor}

\usepackage{ragged2e}
\makeatletter
\long\def\@makecaption#1#2{%
  \par\vskip\abovecaptionskip
  \begingroup\small\rmfamily
    \begingroup\samepage\justifying\parindent\z@
      \let\footnote\@footnotemark@gobble
      \@make@capt@title{#1}{#2}\par
    \endgroup
  \endgroup
  \vskip\belowcaptionskip}
\makeatother

\begin{document}

\title{The Geometric Phase as a Diagnostic for Driven-Dissipative Oscillators}

\author{Zeen Sun}
\affiliation{Centre for Quantum Technologies (CQT), National University of Singapore 117543, Singapore}
\author{Yuan Shen}
\affiliation{Centre for Quantum Technologies (CQT), National University of Singapore 117543, Singapore}
\author{Haitao Ding}
\affiliation{Centre for Quantum Technologies (CQT), National University of Singapore 117543, Singapore}
\affiliation{MajuLab, CNRS-UNS-NUS-NTU International Joint Research Unit, UMI 3654, Singapore}
\author{Yuancheng Zhan$^*$}
\affiliation{Centre for Quantum Technologies (CQT), National University of Singapore 117543, Singapore}
\affiliation{School of Electrical and Electronic Engineering (EEE), Nanyang Technological University, 50 Nanyang Avenue, Singapore 639798}
\author{Leong-Chuan Kwek$^\dagger$}
\affiliation{Centre for Quantum Technologies (CQT), National University of Singapore 117543, Singapore}
\affiliation{MajuLab, CNRS-UNS-NUS-NTU International Joint Research Unit, UMI 3654, Singapore}
\affiliation{School of Electrical and Electronic Engineering (EEE), Nanyang Technological University, 50 Nanyang Avenue, Singapore 639798}
\affiliation{National Institute of Education (NIE),
Nanyang Technological University, 1 Nanyang Walk, Singapore 637616}

\def\kwek#1{\textcolor{red}{#1}}

\begin{abstract}
Driven-dissipative quantum oscillators lock their phase, deform their limit cycles, and undergo dissipative phase transitions, yet these behaviors are read from unrelated quantities defined on the same steady-state density matrix. We show that a single geometric quantity organizes them. Winding the phase of the drive generates a closed loop of nonequilibrium steady states, and because the Liouvillian is covariant under number rotations, the kinematic mixed-state geometric phase of this loop reduces exactly to an eigensystem functional of a single steady state. Under weak driving, it is governed by the same nearest-neighbor coherences that produce phase locking and inherits the Arnold tongue of synchronization. Near the Hopf threshold, it registers the nonperturbative reorganization of the steady-state eigenvectors. And in the squeezing-driven Kerr resonator, it develops distinct signatures at the first and second order dissipative phase transitions. The geometric phase thus
provides a unified and experimentally accessible characterization of steady-state reorganization.

\end{abstract}

\maketitle
\begingroup
\renewcommand{\thefootnote}{\fnsymbol{footnote}}
\footnotetext[1]{Corresponding author:
zhan0530@e.ntu.edu.sg}
\footnotetext[2]{Corresponding author:
cqtklc@nus.edu.sg}
\endgroup

\textit{Introduction---}Driven-dissipative quantum oscillators are paradigmatic nonequilibrium
systems whose steady states encode phase locking, nonlinear response, and
criticality. The quantum Stuart--Landau (QSL) oscillator is a central
example, proposed and recently realized in trapped
ions~\cite{lee2013quantum,walter2014quantum,li2025experimental}, exhibiting
genuinely quantum synchronization effects such as
blockade~\cite{PhysRevLett.117.073601,lorch2017quantum} and a variety of
measures and enhancement
protocols~\cite{PhysRevA.99.043804,PhysRevLett.120.163601,
PhysRevResearch.2.033422,jaseem_generalized_2020,kato2021enhancement,
shen2023enhancing,shen2023fisher_info,shen2023nonlinear,
vaidya2024quantum_sync_and_diss_quantum_sensing}. Related nonlinear
oscillators exhibit relaxation oscillations~\cite{chia2020relaxation},
multistability~\cite{ahmadi2023extreme,hens2015extreme}, and bursting
dynamics~\cite{zhang2023bursting}, from circadian
clocks~\cite{gonze2005spontaneous} to complex
networks~\cite{wu2007synchronization}. Another key platform is the
two-photon-driven Kerr resonator, where parametric
driving~\cite{wilson2010photon}, Kerr nonlinearity, and engineered
dissipation~\cite{leghtas2015confining} stabilize cat
qubits~\cite{Puri2017npj,puri2019prx,Grimm2020nature,
ding2024quantum_osc+kerr_qubit,yu2024stable_kerr_arbitary_cat} and give rise
to dissipative phase transitions
(DPTs)~\cite{bartolo2016exact,PhysRevA.98.042118,beaulieu_observation_2025}.
These examples motivate a unified view of synchronization and criticality as
forms of steady-state reorganization.

Geometric phase (GP) provides a language for global state-space geometry.
Since Berry's formulation for adiabatic pure-state
cycles~\cite{berry1984quantal}, with earlier roots in Pancharatnam's phase
and classical geometric
effects~\cite{pancharatnam1956generalized,hamilton1831third,lloyd1831phenomena},
GP has been extended to mixed states through Uhlmann's
phase~\cite{uhlmann1986parallel} and the interferometric
phase~\cite{sjoqvist2000gp_for_mixed_state_in_interferometry}, and to open
dynamics through the kinematic approach for nonunitary
evolution~\cite{tong2004kinematic} alongside
open-system~\cite{carollo2003gp_in_open_sys},
trajectory-based~\cite{sjoqvist2006geometric},
channel-based~\cite{ericsson2003generalization,kult2008holonomy}, and
adiabatic~\cite{sarandy2005adiabatic} formulations. These tools have found
applications in quantum
information~\cite{sjoqvist2015gp_in_quantum_information,zhang2023geometric},
multipartite
systems~\cite{yi2004effect_of_intersubsystem_on_gp,
niu2010separable_states_and_gp_of_interacting_two_spin}, coherent
states~\cite{chaturvedi1987berry}, and dissipative circuits and
oscillators~\cite{pechal2012geometric,vacanti2012geometric,
chattopadhyay2018finding,viotti2024geometric,nikdel2024coherent}. For
driven-dissipative oscillators, winding the phase of an external drive
naturally generates a closed loop of nonequilibrium steady states,
suggesting that the GP could serve as a probe of the steady-state manifold.

Existing diagnostics of driven-dissipative response remain fragmented. Synchronization is commonly measured by phase coherence, Hopf-critical response by observables such as the average of the annihilation operator, and dissipative phase transitions by order-parameter behavior and Liouvillian gap closing. These quantities are tailored to different phenomena, but they probe different projections of the same steady-state density matrix. A recent study showed numerically that a spin-$1$ limit-cycle oscillator acquires a mixed-state GP with an Arnold-tongue structure~\cite{daniel2023geometric}. Two problems remain open. First, for continuous-variable oscillators the kinematic GP has not been defined through a natural closed loop of nonequilibrium steady states. Second, it is unknown whether the GP carries physical content beyond phase locking, in particular near critical points where no perturbative structure survives. Consequently, the central question becomes whether mixed-state GP can serve as a unifying geometric diagnostic for steady-state reorganization, connecting synchronization and criticality within a single state-space invariant.

\begin{figure}[htbp]
    \makebox[\linewidth][c]{\includegraphics[width=0.8\linewidth]{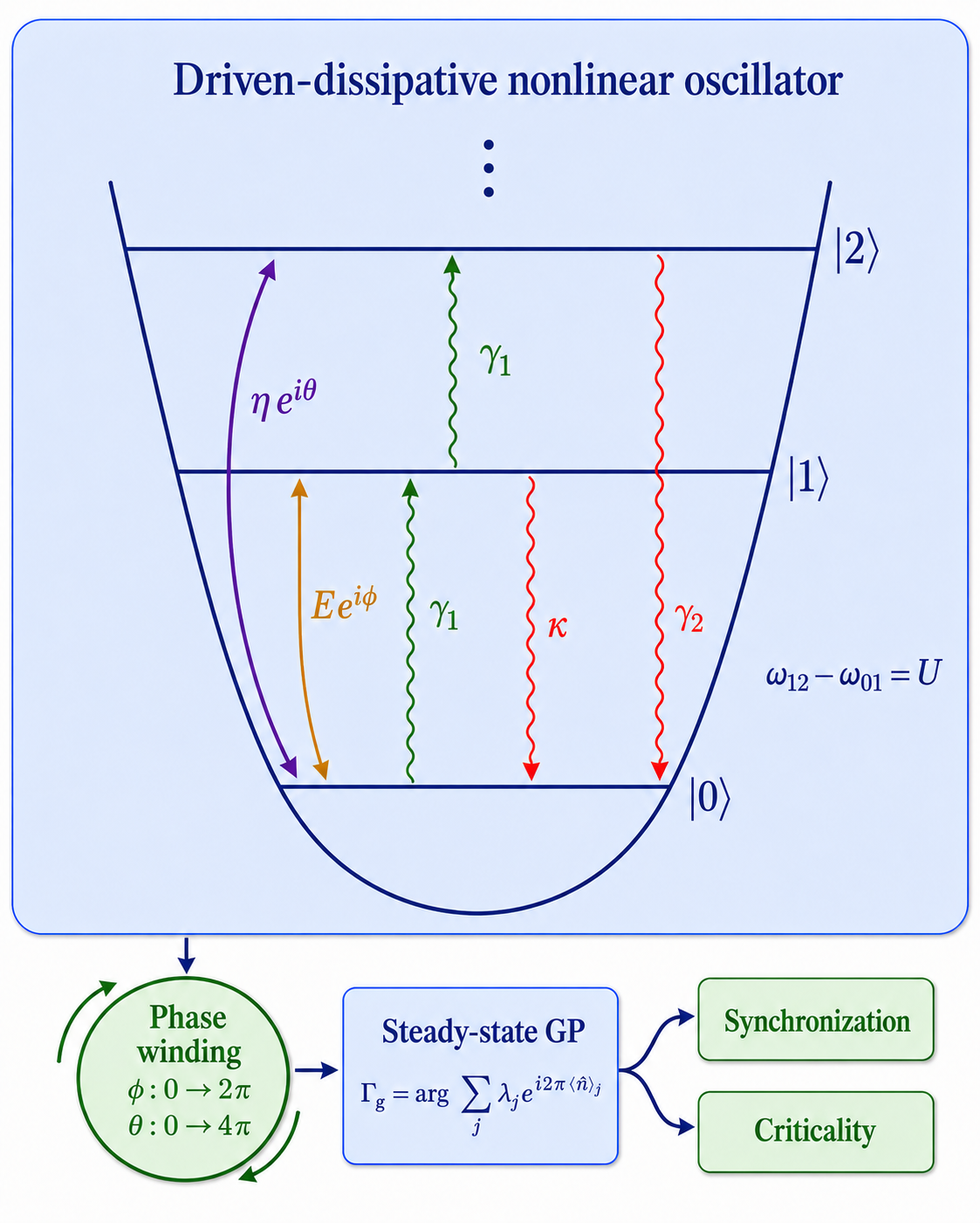}}
    \caption{Schematic of the driven-dissipative nonlinear oscillator \eqref{eqn:master_eq} and phase-winding protocol used to generate the steady-state GP, which probes synchronization and criticality.}
    \label{fig:schematic_plot}
\end{figure}

Here we show that the mixed-state GP generated by phase winding provides a unified diagnostic of steady-state reorganization across synchronization, Hopf-critical response, and DPT. The GP associated with the closed trajectory therefore reduces to an exact eigensystem functional of the steady-state density matrix. In the weak-drive QSL regime, this quantity is controlled by the same coherences that generate phase locking, so that the GP inherits an Arnold-tongue-like resonance structure from the same coherence sector that produces phase locking. Beyond this regime, the GP detects steady-state eigensystem reorganization near the Hopf-critical response of the harmonically driven QSL oscillator. Applying the same construction to the two-photon-driven Kerr resonator, we further show that the GP develops distinct signatures near the boundaries associated with the first- and second-order DPT. Thus, the GP provides a global geometric probe of steady-state reorganization across phase locking, nonlinear critical response, and DPTs as shown in~\cref{fig:schematic_plot}.

\textit{The physical model---}  
In this study, we investigate a driven-dissipative bosonic system, also known as a generalized QSL with anharmonicity. The natural frequency and the harmonic drive frequency are denoted by $\omega_0$ and $\omega$, respectively. For simplicity, we take the squeezing-drive frequency to be $2\omega$. The master equation in the rotating frame is therefore (with $\hbar = 1$): 
\begin{align}\label{eqn:master_eq}
    \dot \rho = & -i[\hat{H},\rho] + \gamma_1 \mathcal{D}[a^\dagger]\rho +\gamma_2\mathcal{D}[a^2]\rho + \kappa\mathcal{D}[a]\rho \nonumber \\
    \hat{H} = & \, \begin{aligned}[t]
        &\Delta a^\dagger a + E (ae^{i\phi}  + a^{\dagger} e^{-i\phi} ) \\
        &+ \eta (a^2 e^{i\theta} + a^{\dagger 2}e^{-i\theta}) + U a^{\dagger2}a^2
    \end{aligned}
\end{align}
where $\mathcal{D}[L]\rho = L\rho L^\dagger - \frac{1}{2}(L^\dagger L \rho + \rho L^\dagger L)$.  Here, $\gamma_1$, $\gamma_2$, and $\kappa$ are the rates of single-photon pumping, nonlinear damping, and linear damping, respectively. The detuning is $\Delta= \omega_0 -\omega$. $E$ and $\eta$ denote the strengths of the harmonic and squeezing drives, respectively. The squeezing term itself acts as a two-photon drive, even in the absence of the harmonic drive \cite{PhysRevLett.120.163601}. Here, $U$ denotes the strength of the Kerr nonlinearity, which is generally unavoidable in experimental implementations. Different limits of \eqref{eqn:master_eq} capture the essential dynamics realized in trapped-ion quantum oscillators and parametrically driven Kerr resonators \cite{li2025experimental,beaulieu_observation_2025}. It is assumed that the squeezing-drive frequency is twice the harmonic-drive frequency in \eqref{eqn:master_eq}.

For the synchronization analysis, we write $\mathcal L=\mathcal L_0+E\mathcal L_{\rm ext}$, where $\mathcal L_{\rm ext}\rho=-i[\hat H_{\rm ext},\rho]$, and treat the external harmonic drive perturbatively. This formalism is useful for studying quantum synchronization where the quantum self-sustained oscillator can be locked to the external drive. In this context, the external drive may be either harmonic or squeezing.

\textit{Kinematic GP---} We use the kinematic approach to obtain the GP of the oscillator~\cite{tong2004kinematic}, which is expressed as follows:
\begin{multline}\label{eqn:GP}
    \gamma[P] = arg\bigg[\sum_{k=1}^{N}\sqrt{p_k(0)p_k(\tau)}\langle\phi_k(0)|\phi_k(\tau)\rangle \\
    \times exp(-\int_{0}^{\tau}\langle\phi_k(t)|\dot\phi_k(t)\rangle dt)\bigg],
\end{multline}
where $p_k(t)$ and $|\phi_k(t)\rangle$ are the $k$-th eigenvalue and eigenstate of the system density matrix $\rho$ at time $t$.

Winding the phase of the drive drags the steady state around the oscillator phase space by a unitary. As shown in the Supplementary Material~\cite{suppmat}, the protocol is just a number-rotation covariance of the steady state:
\begin{equation}\label{eqn:covariance}
\rho_{\rm ss}(\varphi)=R(\varphi)\rho_{\rm ss}(0)R^\dagger(\varphi),\qquad R(\varphi)=e^{-i\varphi \hat n}, 
\end{equation} 
where $\varphi=\phi$ for the harmonic drive, while $\varphi=\theta/2$ for the squeezing drive. Thus a closed phase-space loop corresponds to $\varphi:0\rightarrow2\pi$. If $\rho_{\rm ss}(0)=\sum_j\lambda_j|\psi_j\rangle\langle\psi_j|$, then along the loop the eigenvalues are unchanged and $|\phi_j(t)\rangle=R[\varphi(t)]|\psi_j\rangle$. Substituting this into \eqref{eqn:GP}, and using $e^{-i2\pi\hat n}=1$, gives
\begin{equation}\label{eqn:GP_exact} 
\Gamma_{g}=\arg Z = \arg \sum_j\lambda_j e^{i2\pi\langle\hat n\rangle_j},
\end{equation} 
where $\displaystyle \langle\hat n\rangle_j=\langle\psi_j|\hat n|\psi_j\rangle$. 

This is the central result of this work. We emphasize that the kinematic GP \eqref{eqn:GP} is defined for an arbitrary path of density operators, with no adiabaticity requirement. The obstacle is that evaluating it requires the spectral decomposition of $\rho$ at every point of the loop. The covariance removes this exactly since the eigenvalues are
constant and the eigenvectors are rigidly transported by $R(\varphi)$. The resulting $Z=\sum_j \lambda_j e^{i2\pi\langle\hat n\rangle_j}$ therefore is only a nonlinear functional of the steady state. \eqref{eqn:GP_exact} is sensitive not to the mean photon number but to how the eigenmodes are distributed in number space. Phase winding thereby converts steady-state reorganization into a geometric response, accessible from state reconstruction.

\textit{Perturbative GP and quantum synchronization---} Having obtained the exact GP in \eqref{eqn:GP_exact}, we now ask how it behaves when only the harmonic drive weakly perturbs the QSL. In this regime, the drive-induced off-diagonal coherences provide the usual synchronization response~\cite{PhysRevA.99.043804}, while the GP captures the corresponding geometric response of the steady-state eigensystem. We will show that both quantities are connected through the first-order coherences.

To quantify synchronization, we use the phase coherence. For a density matrix written in the Fock basis, the phase distribution contains Fourier components determined by its off-diagonal elements, $\displaystyle P(\varphi)=1/2\pi\sum_{m,n}\rho_{mn}e^{i(n-m)\varphi}$.
The $q$th-order phase coherence is defined as
\begin{equation}\label{eqn:phase_coherence}
    S_q=\left|\int_0^{2\pi}d\varphi\,P(\varphi)e^{iq\varphi}\right|
    =
    \left|\sum_n \rho_{n+q,n}\right|.
\end{equation}
For a harmonic drive, the relevant synchronization response is the first-order coherence $S_1$, while a squeezing drive naturally probes the two-photon coherence $S_2$, corresponding to $2$-to-$1$ phase locking.

We first consider a purely harmonic drive without anharmonicity, where $\displaystyle\mathcal{L}_{\rm ext}\rho=-i[ae^{i\phi}+a^{\dagger}e^{-i\phi},\rho]$,
and the total Liouvillian is $\mathcal{L}=\mathcal{L}_0+E\mathcal{L}_{\rm ext}$. 
The unperturbed QSL limit cycle is diagonal in the Fock basis, $\displaystyle \rho^{(0)}=\sum p_n |n\rangle\langle n|$~\cite{zybb-vxfz}. Since the harmonic perturbation is off diagonal in this basis, the first-order correction has vanishing diagonal elements and generates only nearest-neighbor coherences,
\begin{equation}
    \rho_{\rm ss}(\phi)\simeq\rho^{(0)}+E\rho^{(1)}(\phi).
\end{equation}
Thus, the first-order phase coherence scales linearly with the drive strength,
\begin{equation}
    S_1=\left|\sum_{n=0}^\infty\rho_{n+1,n}\right| 
    =
    E\left|\sum_{n=0}^\infty x_n\right|+O(E^2),
\end{equation}
where $x_n=\left.\partial_E\rho_{n,n+1}(E)\right|_{E=0}$.

The same coherence susceptibility also determines the perturbative GP. Starting from the analytical expression in \eqref{eqn:GP_exact}, the eigenvalues remain $p_n+O(E^2)$ because the first-order correction is off diagonal, while the eigenvectors acquire $O(E)$ mixtures between neighboring Fock states. Since $\hat n$ is diagonal, the corresponding correction to $\langle\hat n\rangle_n$ starts at order $E^2$. Substituting this perturbative expansion into \eqref{eqn:GP_exact} gives the leading nonvanishing GP
\begin{equation}\label{eqn:perturb_GP}
    \Gamma_g
    =
    2\pi E^2
    \sum_n
    \frac{|x_n|^2}{p_n-p_{n+1}}
    +O(E^4).
\end{equation}
\eqref{eqn:perturb_GP} is the nondegenerate weak-drive expansion of the GP. It shows that the phase coherence and the GP are controlled by the same drive-induced nearest-neighbor coherence: $S_1$ is linear in this coherence, while $\Gamma_g$ is its quadratic imprint. Thus the GP is not an independent synchronization measure, but rather a geometric susceptibility of the perturbed steady-state manifold.

\begin{figure}[htbp]
    \makebox[\linewidth][c]{\includegraphics[width=1.1\linewidth]{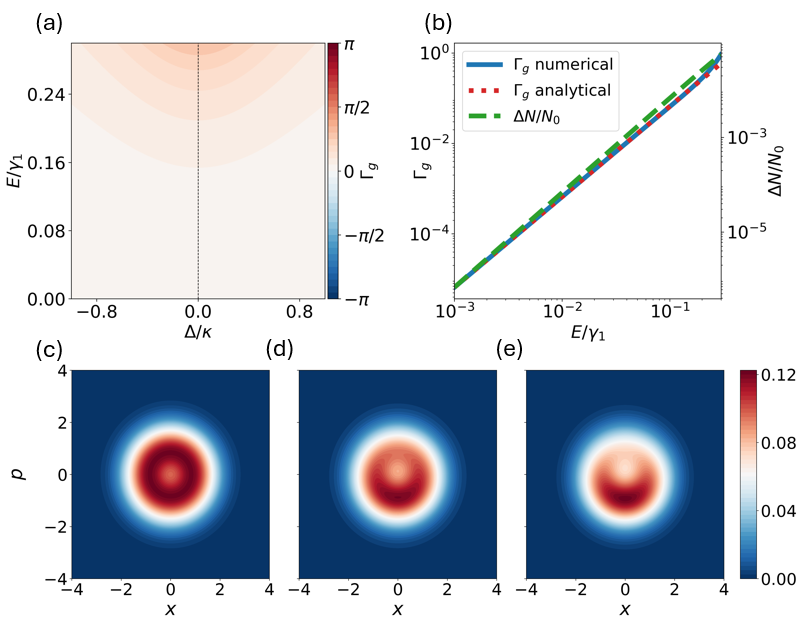}}
    \caption{
    Perturbative GP response. $\gamma_2/\gamma_1 = 3$ and $\kappa/\gamma_1=0.1$
    (a) Arnold tongue in this quantum regime.
    (b) The GP response at zero detuning. Blue solid line indicates the numerical GP, while the orange dotted line is obtained by perturbation. The green dashed line indicates the relative change in the limit-cycle amplitude as the drive strength increases.
    (c)-(e) The Wigner functions at $E/\gamma_1=0$, $0.1$, and $0.2$, respectively.}
    \label{fig:perturb_GP}
\end{figure}

\cref{fig:perturb_GP}(a) shows that the GP forms an Arnold-tongue-like response in the detuning--drive plane, reflecting the resonantly enhanced phase-locking response of the limit cycle, similar to the phase coherence. To check that the system remains in the perturbative regime, we monitor the relative amplitude deformation,
$\Delta N/N_0=(\langle a^\dagger a\rangle_E-\langle a^\dagger a\rangle_0)/\langle a^\dagger a\rangle_0$,
where $N_0=\langle a^\dagger a\rangle_0$. For the weak drives considered, $\Delta N/N_0$ remains small, so the drive mainly locks the phase rather than deforming the limit cycle. In the quantum regime $\gamma_1\leq\gamma_2$, the numerical GP follows the predicted quadratic scaling for $E/\gamma_1\lesssim0.2$, as shown in~\cref{fig:perturb_GP}(b), with $\Delta N/N_0\simeq0.02$.

The quadratic behavior is valid in the strongly damped system even though the system is driven by squeezing. We show this numerically and analytically, especially in the deep quantum regime $\gamma_1 \ll \gamma_2$ where the dynamics can be restricted to a three-level subspace~\cite{suppmat}.
This is because an unperturbed strongly damped QSL can often be truncated to the lowest few Fock levels. However, perturbation theory breaks down when the system is strongly driven, especially in the semiclassical regime. This motivates using the full expression in \eqref{eqn:GP_exact}, which remains well defined beyond the weak-coherence regime and can diagnose more general changes of the driven steady state.

\textit{Exact analysis and critical behavior---}We now consider the regime where the perturbative expansion is no longer sufficient, while the full geometric expression in \eqref{eqn:GP_exact} remains applicable. This occurs near the semiclassical threshold of the QSL oscillator, where $\gamma_2\rightarrow0$ and $\gamma_1$ becomes comparable to $\kappa$. At the classical level, the effective damping $\mu=(\gamma_1-\kappa)/2$ vanishes at the Hopf bifurcation, where the stable fixed point loses stability and a stable limit cycle emerges. In the quantum model, this crossover reorganizes the steady-state density-matrix eigensystem; the GP also probes the full geometric amplitude $Z=\sum_j\lambda_j e^{i2\pi\langle \hat n\rangle_j}$. Thus, the kinematic GP acts as a geometric response of the driven steady-state manifold, diagnosing both weak phase locking and the critical reorganization near the Hopf threshold.

\begin{figure}[htbp]
\centering
    \makebox[\linewidth][c]{\includegraphics[width=1.1\linewidth]{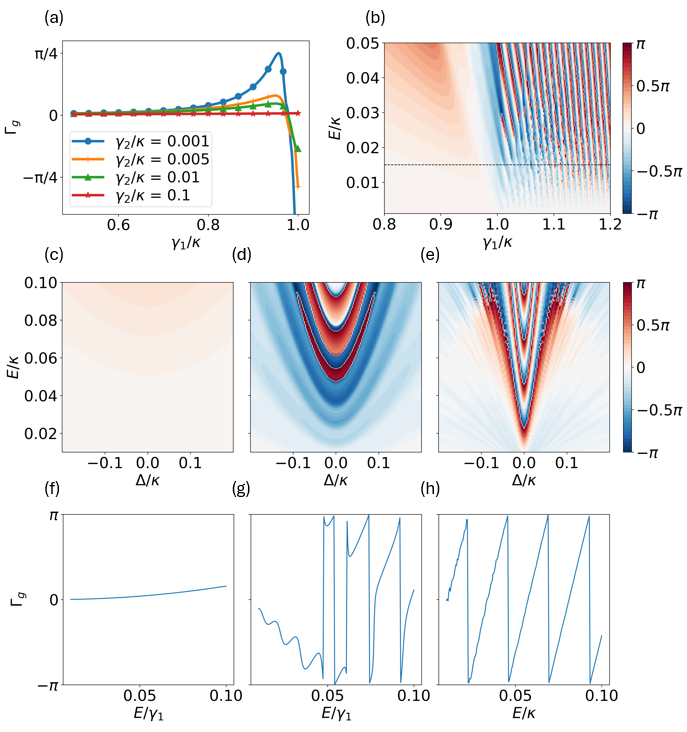}}
    \caption{(a) GP for different values of $\gamma_2/\kappa$ at the same drive strength $E/\kappa = 0.015$. Smaller nonlinear damping produces a sharper change in the GP near the critical point. (b) GP diagram at $\gamma_2/\kappa= 0.005$ and $\Delta=0$. The black dotted line denotes the orange curve from (a). (c)-(e) GP as a function of $E/\kappa$ and $\Delta/\kappa$ at fixed $\gamma_1/\kappa = 0.6, 1, 1.4$, respectively. (f) - (h) Zero-detuning cross sections for $\gamma_1/\kappa = 0.6, 1, 1.4$\hfill}
    \label{fig:phase_diagram}
\end{figure}

We now turn on a weak harmonic drive to probe the geometric response near the Hopf threshold. As shown in~\cref{fig:phase_diagram}(a) and (b), the GP first increases as $\mu\to0^-$ and then drops rapidly near the critical region. The enhancement is stronger for smaller $\gamma_2$, since weaker nonlinear damping broadens the steady-state number distribution and reduces the adjacent population gaps $p_n-p_{n+1}$, thereby enhancing the perturbative response in \eqref{eqn:perturb_GP}.

This perturbative picture breaks down sufficiently close to threshold. At $E=0$, the eigenvectors are Fock states and $\langle\hat n\rangle_j$ is integer, so the GP vanishes. A weak drive only produces small fractional shifts $\langle\hat n\rangle_j=m_j+f_j$ with $f_j\ll1$, giving the quadratic GP. Near the critical region, however, the small population gaps allow the drive to strongly reorganize the eigenvectors of $\rho_{\rm ss}$, so some $f_j$ become $O(1)$. The phasors $\lambda_j e^{i2\pi f_j}$ then rotate substantially, and their sum $Z=\sum_j\lambda_j e^{i2\pi f_j}$ can bend or partially cancel. The sharp GP drop and the subcritical fan therefore reflect nonperturbative phasor interference in the density-matrix eigensystem, while the multiple lobes inside the fan arise from the same drive-induced coherence channel beyond the quadratic regime.

In~\cref{fig:phase_diagram}(c)--(e) and the zero-detuning cuts in~\cref{fig:phase_diagram}(f)--(h), the GP shows three qualitatively different responses. For $\gamma_1/\kappa=0.6$, the system is still in the damped regime, so the weak-drive expansion in \eqref{eqn:perturb_GP} remains valid and the zero-detuning GP is quadratic in $E$. For $\gamma_1/\kappa=1.4$, a limit-cycle amplitude already exists. The drive mainly fixes the phase of this limit cycle, as the GP changes approximately linearly with $E$, giving an almost linear response modulo $2\pi$. The critical response at $\mu=0$ is qualitatively different. The drive strongly reorganizes the dominant eigenvectors, making their fractional number shifts $f_j=\langle\hat n\rangle_j~{\rm mod}~1$ change rapidly with $E$. Since $\operatorname{Im}Z=\sum_j\lambda_j\sin(2\pi f_j)$, different eigenvectors alternately add positive and negative contributions to $\operatorname{Im}Z$. This produces the oscillatory descent of the GP. 

We can conclude that perturbative GP should be viewed as a small coherence expansion of the phasor away from the critical point. Near the Hopf threshold, however, the steady-state eigensystem is strongly reorganized: the fractional photon-number parts of the eigenvectors become nonperturbative, and the full phasor $Z$ must be used. The kinematic GP therefore provides a geometric diagnostic of the critical reorganization of the steady state.

\textit{GP and DPT---}We now set $E=0$ and retain the Kerr nonlinearity and squeezing drive. The latter reduces the continuous $U(1)$ number-rotation symmetry to parity, $\Pi=e^{i\pi\hat n}$, with $\Pi\hat a\Pi^\dagger=-\hat a$. The resulting $\mathbb{Z}_2$-symmetric Kerr resonator supports a bright lobe bounded by a second-order DPT at lower detuning and a first-order DPT at higher detuning. We ask whether the same geometric amplitude resolves the distinct steady-state reorganizations underlying these transitions.

To characterize the local response without introducing a phase-unwrapping convention, we define
\begin{equation}
\chi_{\Gamma_{\rm g}}
=
\operatorname{Im}\left(\frac{\partial_x Z}{Z}\right),
\qquad
x=\Delta/|U|,
\label{eq:gp_response}
\end{equation}
which equals $\partial_x\arg Z$ wherever $Z\neq0$. Differentiating $Z$ separates the response as
$\chi_{\Gamma_{\rm g}}=\chi_\lambda+\chi_n$, where
$\chi_\lambda$ describes the redistribution of the density-matrix eigenvalues, while $\chi_n$ describes changes in the number-space structure of the corresponding eigenvectors. Explicit expressions are given in the Supplementary Material \cite{suppmat}. Although $|Z|$ remains nonzero at both transition boundaries, partial cancellation among the eigenmode phasors inside the bright phase can reduce $|Z|$ and enhance the local response.

\begin{figure}[htbp]
\centering
\makebox[\linewidth][c]{%
\includegraphics[width=1.1\linewidth]{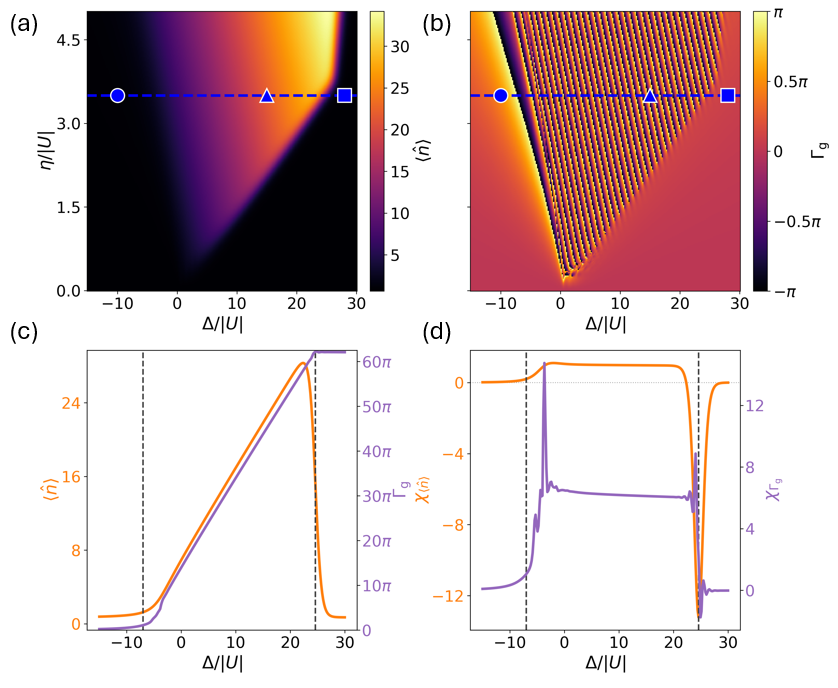}}
\caption{
Geometric response of the squeezing-driven Kerr resonator for
$U<0$ and $\gamma_1=\gamma_2=\kappa=0.1|U|$.
(a) Steady-state photon number $\langle\hat n\rangle$ as a function of
the rescaled squeezing strength $\eta/|U|$ and detuning
$x=\Delta/|U|$. The dotted line marks the cut at $\eta/|U|=3.5$.
The circle, triangle, and square indicate representative
low-detuning vacuum-like, bright, and high-detuning vacuum-like
states, respectively.
(b) GP for the same parameters.
(c) Cross sections of $\langle\hat n\rangle$ and
$\Gamma_{\rm g}/\pi$ at $\eta/|U|=3.5$. The GP is unwrapped only as a
visual guide. The vertical dashed lines mark the independently
identified second- and first-order boundaries.
(d) Local responses
$\chi_{\langle n\rangle}=\partial_x\langle\hat n\rangle$
and
$\chi_{\Gamma_{\rm g}}=\partial_x\Gamma_{\rm g}$.
The left and right windows show the neighborhoods of the second- and
first-order boundaries, respectively, using separate vertical scales.
}
\label{fig:GP_DPT}
\end{figure}

At the lower boundary, $\Gamma_{\rm g}$ evolves continuously, and the boundary-local response is dominated by $\chi_n$, whereas $\chi_\lambda$ remains comparatively small. The GP therefore arises primarily from the continuous deformation of the steady-state eigenvectors, which rotates the corresponding phasors $\lambda_j e^{i2\pi n_j}$.

The largest magnitude of $\chi_{\Gamma_{\rm g}}$ on the negative-detuning side occurs farther inside the bright phase rather than exactly at the transition boundary. There, partial cancellation among the dominant eigenmode phasors reduces $|Z|$ and amplifies the phase response through
$\chi_{\Gamma_{\rm g}}$.
This interior feature therefore reflects geometric interference within the reorganized bright state rather than a displacement of the second-order boundary.

At the upper boundary, the steady state changes rapidly from the bright regime to the vacuum-like regime. The associated redistribution of dominant eigenvalues of the density matrix produces a nonnegligible $\chi_\lambda$, while the simultaneous reorganization of their number-space structure gives a substantial $\chi_n$. These contributions may reinforce or partially cancel one another, explaining why the strongest GP response can be slightly displaced from the maximum photon-number slope. This also explains the interference near the upper boundary for different $\eta$ and $\Delta$.

The two boundaries thus produce distinct geometric signatures: the lower boundary is governed mainly by continuous eigenvector deformation, whereas the upper boundary additionally involves rapid eigenvalue redistribution. The GP therefore probes the reorganization of the complete steady-state eigensystem rather than acting as a conventional order parameter.

\textit{Conclusion---}We have shown that the mixed-state geometric phase reveals a state-space structure underlying the response of bosonic driven-dissipative oscillators. In the QSL oscillator, the geometric response reduces in the weak-drive limit to the same coherence physics responsible for synchronization, but it is not confined to phase locking. When the steady state is strongly deformed, or when the squeezing-driven Kerr resonator approaches dissipative criticality, the geometric phase instead tracks the reorganization of the mixed-state eigensystem. We also providesimple interferometric methods for experimentally determining the GP in the Supplementary Materials \cite{suppmat}.

This establishes a direct link between GP, synchronization, and dissipative criticality as distinct manifestations of steady-state reorganization. The GP therefore provides a compact characterization of the mixed-state eigensystem that is complementary to individual observables such as phase coherence and photon number. Looking ahead, extending the mixed-state GP to interacting many-body open systems may provide a geometric perspective on collective nonequilibrium dynamics and dissipative phase structure \cite{diehl2008quantum}. More broadly, advances in quantum simulation and programmable photonic processors—from large-scale boson sampling to coherent multistep evolution and probes of chaotic dynamics—are making increasingly complex quantum dynamics experimentally accessible \cite{madsen2022quantum,zhan2026loop,zhan2026boson,georgescu2014quantum}, opening a possible route for exploring such geometric diagnostics beyond few-mode settings.

\bibliographystyle{unsrt} 
\bibliography{ref}

\clearpage 

\setcounter{equation}{0}
\setcounter{figure}{0}
\setcounter{table}{0}
\makeatletter
\renewcommand{\theequation}{S\arabic{equation}}
\renewcommand{\thefigure}{S\arabic{figure}}

\end{document}


\title{Supplementary Material for ``The Geometric Phase as a Diagnostic for Driven-Dissipative Oscillators''}

\author{Zeen Sun}
\affiliation{Centre for Quantum Technologies (CQT), National University of Singapore 117543, Singapore}
\author{Yuan Shen}
\affiliation{Centre for Quantum Technologies (CQT), National University of Singapore 117543, Singapore}
\author{Haitao Ding}
\affiliation{Centre for Quantum Technologies (CQT), National University of Singapore 117543, Singapore}
\affiliation{MajuLab, CNRS-UNS-NUS-NTU International Joint Research Unit, UMI 3654, Singapore}
\author{Yuancheng Zhan\textsuperscript{*}}
\affiliation{Centre for Quantum Technologies (CQT), National University of Singapore 117543, Singapore}
\affiliation{School of Electrical and Electronic Engineering (EEE), Nanyang Technological University, 50 Nanyang Avenue, Singapore 639798}
\author{Leong-Chuan Kwek\textsuperscript{*}}
\affiliation{Centre for Quantum Technologies (CQT), National University of Singapore 117543, Singapore}
\affiliation{MajuLab, CNRS-UNS-NUS-NTU International Joint Research Unit, UMI 3654, Singapore}
\affiliation{School of Electrical and Electronic Engineering (EEE), Nanyang Technological University, 50 Nanyang Avenue, Singapore 639798}
\affiliation{National Institute of Education (NIE),
Nanyang Technological University, 1 Nanyang Walk, Singapore 637616}

\maketitle
\begingroup
\renewcommand{\thefootnote}{\fnsymbol{footnote}}
\footnotetext[1]{Corresponding author:
zhan0530@e.ntu.edu.sg}
\footnotetext[1]{Corresponding author:
cqtklc@nus.edu.sg}
\endgroup

\setcounter{equation}{0}
\setcounter{figure}{0}
\setcounter{table}{0}
\setcounter{page}{1}

\renewcommand{\theequation}{S\arabic{equation}}
\renewcommand{\thefigure}{S\arabic{figure}}
\renewcommand{\thetable}{S\arabic{table}}
\setcounter{section}{0}
\setcounter{secnumdepth}{1}
\renewcommand{\thesection}{\Roman{section}}
\renewcommand{\thesubsection}{\thesection.\arabic{subsection}}

\section{Phase rotation and geometric phase}\label{sec:phase_rotation}

In this section, we first show that winding the phase of the drive generates a family of steady states related by a number rotation. We then substitute this covariant family into the kinematic mixed-state GP of Tong \textit{et al.}~\cite{Tong2004} and reduce the complete loop to the eigensystem of one steady-state density matrix. A related connection between GP and quantum synchronization was previously obtained for a spin-1 limit-cycle oscillator~\cite{Daniel2023}; here the construction is applied directly to the continuous-variable oscillators considered in the main text.

The phase winding of the harmonic drive can be understood as a number rotation of the oscillator. We define $R(\phi)=e^{-i\phi\hat n}$, with $\hat n=a^\dagger a$. Under this transformation,
\begin{equation}
R(\phi)aR^\dagger(\phi)=e^{i\phi}a,
\qquad
R(\phi)a^\dagger R^\dagger(\phi)=e^{-i\phi}a^\dagger.
\end{equation}
Therefore, rotating the drive phase is equivalent to applying the number rotation to the zero-phase Hamiltonian,
\begin{equation}
H_\phi=R(\phi)H_0R^\dagger(\phi).
\end{equation}
The number and Kerr terms are invariant under this transformation, while the harmonic-drive term acquires the phase $\phi$.

For the dissipative terms, the number rotation only introduces a phase factor in each jump operator,
\begin{equation}
a\rightarrow e^{i\phi}a,\qquad
a^\dagger\rightarrow e^{-i\phi}a^\dagger,\qquad
a^2\rightarrow e^{i2\phi}a^2.
\end{equation}
Such phase factors do not affect a Lindblad dissipator, since
\begin{equation}
\mathcal D[e^{i\chi}O]\rho=\mathcal D[O]\rho.
\end{equation}
Hence, the dissipative processes are invariant under the number rotation, and the Liouvillian satisfies the covariance relation
\begin{equation}
    \mathcal{L}_{\phi}
    \left[
        R(\phi)\rho R^\dagger(\phi)
    \right]
    =
    R(\phi)
    \mathcal{L}_{0}[\rho]
    R^\dagger(\phi).
    \label{eq:Liouvillian_covariance}
\end{equation}
If the steady state is unique and satisfies
$\mathcal{L}_0[\rho_{\rm ss}(0)]=0$, \eqref{eq:Liouvillian_covariance} implies
\begin{equation}
    \rho_{\rm ss}(\phi)
    =
    R(\phi)\rho_{\rm ss}(0)R^\dagger(\phi).
    \label{eq:ss_rotation}
\end{equation}
Thus, winding the phase of the harmonic drive rotates the steady state in phase space without changing its spectrum.

We now consider a closed trajectory $\phi:0\rightarrow2\pi$. Let the spectral decomposition of the initial steady state be
\begin{equation}
    \rho_{\rm ss}(0)
    =
    \sum_j\lambda_j|\psi_j\rangle\langle\psi_j|,
\end{equation}
where the nonzero eigenvalues are assumed to be nondegenerate. From \eqref{eq:ss_rotation}, the instantaneous eigenvectors are
$|\psi_j(\phi)\rangle=R(\phi)|\psi_j\rangle$, while $\lambda_j$ is independent of $\phi$.

For a path of density operators, the kinematic mixed-state geometric phase (GP) is~\cite{Tong2004}
\begin{equation}
    \Gamma_{\rm g}[\mathcal{P}]
    =
    \arg
    \sum_j
    \sqrt{\lambda_j(0)\lambda_j(\tau)}
    \langle\psi_j(0)|\psi_j(\tau)\rangle
    \exp\left[
        -\int_0^\tau
        \langle\psi_j(t)|\dot{\psi}_j(t)\rangle dt
    \right].
    \label{eq:Tong_GP}
\end{equation}
For the phase-winding loop, the eigenvalues are constant and
\begin{equation}
    \langle\psi_j(0)|\psi_j(\tau)\rangle
    =
    \langle\psi_j|e^{-i2\pi\hat n}|\psi_j\rangle
    =1,
\end{equation}
because $\hat n$ has integer eigenvalues. Moreover,
$\langle\psi_j(t)|\dot{\psi}_j(t)\rangle
=-i\dot{\phi}(t)\langle\psi_j|\hat n|\psi_j\rangle$. Hence
\begin{equation}
    -\int_0^\tau
    \langle\psi_j(t)|\dot{\psi}_j(t)\rangle dt
    =
    i2\pi\langle\hat n\rangle_j,
    \qquad
    \langle\hat n\rangle_j
    =
    \langle\psi_j|\hat n|\psi_j\rangle .
\end{equation}
Substituting this into \eqref{eq:Tong_GP}, we obtain
\begin{equation}
    \Gamma_{\rm g}
    =
    \arg Z,
    \qquad
    Z
    =
    \sum_j
    \lambda_j
    e^{i2\pi\langle\hat n\rangle_j}.
    \label{eq:GP_exact}
\end{equation}
\eqref{eq:GP_exact} is the central reduction used throughout this work. The GP of the closed steady-state trajectory is obtained from the eigensystem of one steady-state density matrix. It is sensitive not simply to the mean photon number, but to how the eigenmodes of the density matrix are distributed in number space.

The kinematic definition in \eqref{eq:Tong_GP} does not require an adiabatic path. The analytical evaluation in \eqref{eq:GP_exact} only uses the steady-state family generated by the covariance relation.

For a pure two-photon drive, the covariance angle is $\phi=\theta/2$, so that
\begin{equation}
    \rho_{\rm ss}(\theta)
    =
    R(\theta/2)\rho_{\rm ss}(0)R^\dagger(\theta/2).
\end{equation}
A $2\pi$ winding of the squeezing phase gives
$R(\pi)=\Pi=e^{-i\pi\hat n}$. For a parity-symmetric steady state with nondegenerate eigenvalues, the eigenvectors may be chosen as parity eigenstates,
$\Pi|\psi_j\rangle=\pi_j|\psi_j\rangle$, where $\pi_j=\pm1$. The corresponding geometric amplitude is
\begin{equation}
    Z_{2\pi}^{(\theta)}
    =
    \sum_j
    \lambda_j\pi_j
    e^{i\pi\langle\hat n\rangle_j}.
    \label{eq:GP_squeeze_2pi}
\end{equation}
Thus, the closed $2\pi$ squeezing-phase loop contains an additional parity endpoint factor. To use the same geometric functional as for harmonic driving, we take
$\theta:0\rightarrow4\pi$, for which $R(2\pi)=\mathbbm{1}$ and
\begin{equation}
    Z_{4\pi}^{(\theta)}
    =
    \sum_j
    \lambda_j
    e^{i2\pi\langle\hat n\rangle_j}.
    \label{eq:GP_squeeze_4pi}
\end{equation}
The $4\pi$ protocol is therefore a full phase-space rotation and is the convention used for the squeezing-driven results in the main text.

\section{Weak-drive expansion and the deep quantum regime}\label{sec:weak_drive}

In this section, we first derive the weak-drive expansion and show that the GP and phase coherence are controlled by the same drive-induced coherence sector. We then specialize to the deep quantum regime, where strong nonlinear damping confines the oscillator to the lowest Fock levels and gives a compact nonperturbative expression. The role of coherences in quantum synchronization and the few-level description of the quantum van der Pol oscillator have been discussed in~\cite{Koppenhofer2019,Mok2020}, while the two-photon synchronization response was introduced in~\cite{Sonar2018}.

We first expand \eqref{eq:GP_exact} for a weak harmonic drive. At $E=0$, the steady state is phase symmetric and diagonal in the Fock basis,
\begin{equation}
    \rho^{(0)}
    =
    \sum_n p_n|n\rangle\langle n|,
    \label{eq:rho0_fock}
\end{equation}
where the populations $p_n$ are assumed to be nondegenerate. We write the Liouvillian as
$\mathcal{L}=\mathcal{L}_0+E\mathcal{L}_{\rm ext}$ and expand
\begin{equation}
    \rho_{\rm ss}(E)
    =
    \rho^{(0)}
    +E\rho^{(1)}
    +O(E^2).
    \label{eq:rho_perturbation}
\end{equation}
On the traceless subspace, the first-order correction may be written as
\begin{equation}
    \rho^{(1)}
    =
    -\mathcal{L}_0^{+}\mathcal{L}_{\rm ext}\rho^{(0)},
    \label{eq:liouvillian_response}
\end{equation}
where $\mathcal{L}_0^{+}$ denotes the pseudoinverse of the unperturbed Liouvillian. Since the harmonic perturbation changes the Fock number by one, the first-order density matrix contains nearest-neighbor coherences,
\begin{equation}
    \rho^{(1)}
    =
    \sum_n
    \left(
        x_n|n\rangle\langle n+1|
        +x_n^*|n+1\rangle\langle n|
    \right),
    \label{eq:harmonic_first_order_density}
\end{equation}
where $x_n=\partial_E\rho_{n,n+1}(E)|_{E=0}$. The same coherences give the leading phase-locking response,
\begin{equation}
    S_1
    =
    \left|\sum_n\rho_{n+1,n}\right|
    =
    E\left|\sum_n x_n\right|+O(E^2).
    \label{eq:S1_perturbative_SM}
\end{equation}

Because \eqref{eq:harmonic_first_order_density} is off diagonal in the Fock basis, the density-matrix eigenvalues are unchanged at first order,
\begin{equation}
    \lambda_n
    =
    p_n+O(E^2).
\end{equation}
The eigenvectors receive the first-order corrections
\begin{equation}
    |\psi_n\rangle
    =
    |n\rangle
    +E\left[
        \frac{x_{n-1}}{p_n-p_{n-1}}|n-1\rangle
        +\frac{x_n^*}{p_n-p_{n+1}}|n+1\rangle
    \right]
    +O(E^2),
    \label{eq:harmonic_eigenvector_perturbation}
\end{equation}
where boundary terms such as $x_{-1}$ are taken to vanish. Since $\hat n$ is diagonal in the Fock basis, the first correction to its expectation value is quadratic,
\begin{equation}
    \langle\hat n\rangle_n
    =
    n+E^2\delta n_n+O(E^4),
\end{equation}
with
\begin{equation}
    \delta n_n
    =
    \frac{|x_n|^2}{(p_n-p_{n+1})^2}
    -
    \frac{|x_{n-1}|^2}{(p_n-p_{n-1})^2}.
    \label{eq:harmonic_number_shift}
\end{equation}
Here the absence of odd powers follows from the number-rotation symmetry: changing $E\rightarrow-E$ is equivalent to shifting the harmonic-drive phase by $\pi$, which does not change the closed-loop GP.

Substituting \eqref{eq:harmonic_number_shift} into \eqref{eq:GP_exact}, and using $e^{i2\pi n}=1$, gives
\begin{equation}
    Z
    =
    1+i2\pi E^2\sum_n p_n\delta n_n+O(E^4).
\end{equation}
The weighted number shift can be rearranged as
\begin{align}
    \sum_n p_n\delta n_n
    &=
    \sum_n p_n
    \left[
        \frac{|x_n|^2}{(p_n-p_{n+1})^2}
        -
        \frac{|x_{n-1}|^2}{(p_n-p_{n-1})^2}
    \right] \nonumber\\
    &=
    \sum_n
    \frac{|x_n|^2}{p_n-p_{n+1}}.
\end{align}
Therefore, the leading nonvanishing GP is
\begin{equation}
    \Gamma_{\rm g}
    =
    2\pi E^2
    \sum_n
    \frac{|x_n|^2}{p_n-p_{n+1}}
    +O(E^4).
    \label{eq:perturb_GP_harmonic}
\end{equation}
Thus, $S_1$ is linear in the nearest-neighbor coherence, while the GP is its quadratic imprint on the steady-state eigensystem. The GP is therefore not an independent synchronization measure, but a geometric susceptibility of the same perturbed steady-state manifold.

The same structure applies to a weak squeezing drive. The leading correction changes the Fock number by two,
\begin{equation}
    \rho^{(1)}_{\rm sq}
    =
    \sum_n
    \left(
        y_n|n\rangle\langle n+2|
        +y_n^*|n+2\rangle\langle n|
    \right),
    \label{eq:squeezing_first_order_density}
\end{equation}
where $y_n=\partial_\eta\rho_{n,n+2}(\eta)|_{\eta=0}$. The corresponding two-photon phase coherence is
\begin{equation}
    S_2
    =
    \left|\sum_n\rho_{n+2,n}\right|
    =
    \eta\left|\sum_n y_n\right|+O(\eta^2).
    \label{eq:S2_perturbative_SM}
\end{equation}
For the full phase-space rotation $\theta:0\rightarrow4\pi$, the perturbative GP becomes
\begin{equation}
    \Gamma_{\rm g}^{\rm sq}
    =
    4\pi\eta^2
    \sum_n
    \frac{|y_n|^2}{p_n-p_{n+2}}
    +O(\eta^4).
    \label{eq:perturb_GP_squeezing}
\end{equation}
Changing $\eta\rightarrow-\eta$ is equivalent to a number rotation by $\pi/2$, so the closed-loop GP is again an even function of the drive strength. \eqref{eq:perturb_GP_squeezing} shows that the harmonic drive probes nearest-neighbor population gaps $p_n-p_{n+1}$, while the squeezing drive probes next-nearest-neighbor gaps $p_n-p_{n+2}$.

The perturbative formulas above identify the relevant coherence channel, but they become unreliable when a population difference in the denominator becomes small or when the drive appreciably changes the populations. To keep a compact expression beyond this expansion, we now use the few-level structure of the strongly nonlinear oscillator.

\begin{figure}
    \centering
    \includegraphics[width=0.94\linewidth]{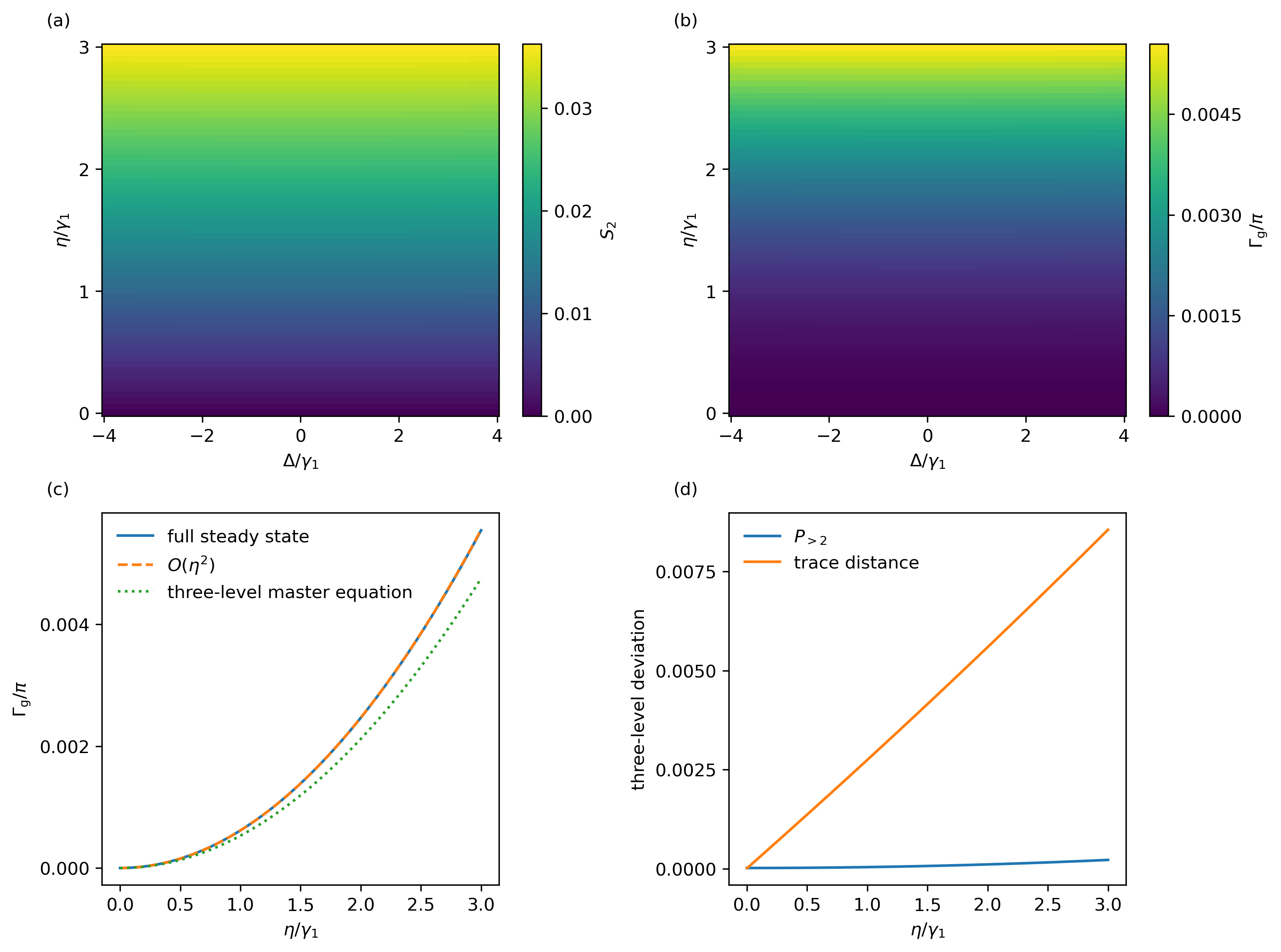}
    \caption{Two-photon coherence and geometric response of the squeezing-driven oscillator in the deep quantum regime. Parameters are $\gamma_2/\gamma_1=100$, $\kappa/\gamma_1=0.1$, and $E=U=0$. (a) Two-photon phase coherence $S_2$ over the extended detuning--squeezing plane. (b) GP for the full $4\pi$ squeezing-phase winding over the same parameter range. (c) Zero-detuning weak-drive cut comparing the full result, \eqref{eq:perturb_GP_squeezing}, and an independent three-level master equation. (d) Population leakage above $|2\rangle$ and trace distance from the normalized three-level projection.}
    \label{fig:squeezing_deep_quantum}
\end{figure}

\medskip
\noindent\textit{Deep quantum regime.---}
We refer to the regime $\gamma_2\gg\gamma_1,\kappa, E,\eta$ as the deep quantum regime. In this limit, nonlinear damping suppresses the population of high Fock states and the steady state can be approximated by the lowest three levels~\cite{Mok2020,Koppenhofer2019}. The purpose of this truncation is not to replace the exact result \eqref{eq:GP_exact}, but to evaluate it in a transparent eigensystem where the relation between coherence and GP can be seen explicitly. For harmonic driving, we use
\begin{equation}
    \rho_{\rm ss}^{(1)}
    \simeq
    \begin{pmatrix}
        \rho_{00} & \rho_{01} & 0\\
        \rho_{10} & \rho_{11} & 0\\
        0 & 0 & \rho_{22}
    \end{pmatrix}.
    \label{eq:rho_three_level_harmonic}
\end{equation}
We define
\begin{equation}
    \Sigma_{01}=\rho_{00}+\rho_{11},
    \qquad
    d_{01}=\rho_{00}-\rho_{11},
    \qquad
    R_{01}=\sqrt{d_{01}^2+4|\rho_{01}|^2}.
\end{equation}
The two eigenvalues of the $|0\rangle$--$|1\rangle$ block are
\begin{equation}
    \lambda_{\pm}^{(01)}
    =
    \frac{\Sigma_{01}\pm R_{01}}{2},
\end{equation}
and their number expectations are
\begin{equation}
    n_{\pm}^{(01)}
    =
    \frac{1}{2}
    \left(1\mp\frac{d_{01}}{R_{01}}\right).
    \label{eq:n_three_level_harmonic}
\end{equation}
The third eigenstate is $|2\rangle$, with eigenvalue $\rho_{22}$ and number expectation $2$. Substitution into \eqref{eq:GP_exact} gives the exact geometric amplitude within the three-level ansatz,
\begin{align}
    Z_{3}^{(1)}
    &=
    \lambda_{+}^{(01)}e^{i2\pi n_{+}^{(01)}}
    +\lambda_{-}^{(01)}e^{i2\pi n_{-}^{(01)}}
    +\rho_{22} \nonumber\\
    &=
    \rho_{22}
    -\Sigma_{01}\cos\left(\pi\frac{d_{01}}{R_{01}}\right)
    +iR_{01}\sin\left(\pi\frac{d_{01}}{R_{01}}\right).
    \label{eq:Z_three_level_harmonic}
\end{align}
For $|\rho_{01}|\ll|d_{01}|$, this becomes
\begin{equation}
    Z_{3}^{(1)}
    =
    1+i2\pi\frac{|\rho_{01}|^2}{d_{01}}
    +O(|\rho_{01}|^4),
\end{equation}
which reproduces \eqref{eq:perturb_GP_harmonic} when the lowest coherence dominates. In this limit, $S_1\simeq|\rho_{01}|$, while the GP is quadratic in the same coherence. When $d_{01}$ becomes small, the perturbative expression is no longer sufficient, but \eqref{eq:Z_three_level_harmonic} remains finite and describes the full mixing of the two density-matrix eigenvectors.

For a pure squeezing drive, the leading coherence lies in the $|0\rangle$--$|2\rangle$ block, and the appropriate three-level ansatz is
\begin{equation}
    \rho_{\rm ss}^{(2)}
    \simeq
    \begin{pmatrix}
        \rho_{00} & 0 & \rho_{02}\\
        0 & \rho_{11} & 0\\
        \rho_{20} & 0 & \rho_{22}
    \end{pmatrix}.
    \label{eq:rho_three_level_squeezing}
\end{equation}
We define
\begin{equation}
    \Sigma_{02}=\rho_{00}+\rho_{22},
    \qquad
    d_{02}=\rho_{00}-\rho_{22},
    \qquad
    R_{02}=\sqrt{d_{02}^2+4|\rho_{02}|^2}.
\end{equation}
The eigenvalues of the even-parity block are
\begin{equation}
    \lambda_{\pm}^{(02)}
    =
    \frac{\Sigma_{02}\pm R_{02}}{2},
\end{equation}
with number expectations
\begin{equation}
    n_{\pm}^{(02)}
    =
    1\mp\frac{d_{02}}{R_{02}}.
    \label{eq:n_three_level_squeezing}
\end{equation}
The odd state $|1\rangle$ has eigenvalue $\rho_{11}$ and number expectation $1$. For the $4\pi$ squeezing-phase loop, the three-level geometric amplitude is therefore
\begin{align}
    Z_{3}^{(2)}
    &=
    \lambda_{+}^{(02)}e^{i2\pi n_{+}^{(02)}}
    +\lambda_{-}^{(02)}e^{i2\pi n_{-}^{(02)}}
    +\rho_{11} \nonumber\\
    &=
    \rho_{11}
    +\Sigma_{02}\cos\left(2\pi\frac{d_{02}}{R_{02}}\right)
    -iR_{02}\sin\left(2\pi\frac{d_{02}}{R_{02}}\right).
    \label{eq:Z_three_level_squeezing}
\end{align}
For $|\rho_{02}|\ll|d_{02}|$,
\begin{equation}
    Z_{3}^{(2)}
    =
    1+i4\pi\frac{|\rho_{02}|^2}{d_{02}}
    +O(|\rho_{02}|^4),
\end{equation}
which is the lowest-level form of \eqref{eq:perturb_GP_squeezing}. Thus, $S_2\simeq|\rho_{02}|$ and the squeezing-driven GP is the quadratic geometric response of the same two-photon coherence.

We numerically verify that the GP in the deep quantum regime has the same resonant dependence on the detuning and squeezing strength under weak squeezing drive as shown in \cref{fig:squeezing_deep_quantum} (a) and (b). Along the zero-detuning cut, $S_2$ is linear in $\eta$, while $\Gamma_{g}$ is quadratic, as predicted by \eqref{eq:S2_perturbative_SM} and \eqref{eq:perturb_GP_squeezing}. The full numerical result follows the weak-drive expression throughout the displayed range. The independent three-level master equation reproduces the same scaling, with a small difference in magnitude caused by the coherences neglected by the strict three-level projection.

The three-level approximation is controlled by the population leakage outside the lowest Fock levels and by the neglected coherences. It is expected to work best for $\gamma_2\gg\gamma_1,\kappa,E,\eta$, and to break down when the drive strongly deforms the limit cycle or populates higher levels.

\section{Geometric amplitude and critical response}\label{sec:critical_response}

The weak-drive expansion describes phase locking while the relevant population gaps remain well separated. We now use the full geometric amplitude $Z$ to analyze the two regimes in which this assumption fails: the Hopf-critical response of the QSL oscillator and the dissipative phase transitions of the two-photon-driven Kerr resonator. The critical response of a quantum VdP oscillator was studied in Ref.~\cite{Dutta2019}, while the steady-state and Liouvillian structures of Kerr DPTs were developed in Refs.~\cite{Bartolo2016,Minganti2018} and observed experimentally in Ref.~\cite{Beaulieu2025}.

The GP is the argument of the complex geometric amplitude
\begin{equation}
    Z(x)
    =
    \sum_j
    \lambda_j(x)e^{i2\pi n_j(x)},
    \qquad
    n_j(x)
    =
    \langle\psi_j(x)|\hat n|\psi_j(x)\rangle,
    \label{eq:Z_definition_SM}
\end{equation}
where $x$ denotes a control parameter. Its local response is evaluated directly from $Z$,
\begin{equation}
    \chi_{\Gamma_{\rm g}}
    =
    \partial_x\arg Z
    =
    \operatorname{Im}\left(\frac{\partial_x Z}{Z}\right)
    =
    \frac{\operatorname{Im}\left[Z^*(\partial_x Z)\right]}{|Z|^2},
    \label{eq:chi_gp_SM}
\end{equation}
provided $Z\neq0$. This definition does not depend on an unwrapping convention for the displayed phase. Differentiating \eqref{eq:Z_definition_SM} gives
\begin{equation}
    \partial_x Z
    =
    \sum_j
    e^{i2\pi n_j}
    \left[
        \partial_x\lambda_j
        +i2\pi\lambda_j\partial_x n_j
    \right].
    \label{eq:dZ_decomposition}
\end{equation}
It is useful to separate the two contributions,
\begin{align}
    \chi_{\lambda}
    &=
    \operatorname{Im}\left[
        \frac{1}{Z}
        \sum_j e^{i2\pi n_j}\partial_x\lambda_j
    \right],
    \label{eq:chi_lambda_SM}\\
    \chi_{n}
    &=
    \operatorname{Im}\left[
        \frac{i2\pi}{Z}
        \sum_j\lambda_j e^{i2\pi n_j}\partial_x n_j
    \right],
    \label{eq:chi_n_SM}
\end{align}
so that $\chi_{\Gamma_{\rm g}}=\chi_{\lambda}+\chi_n$. The first term describes redistribution of statistical weight among the density-matrix eigenmodes, while the second describes changes in their number-space structure.

$|Z|$ quantifies cancellation among the eigenmode phasors
\begin{equation}
    z_j
    =
    \lambda_j e^{i2\pi n_j},
    \qquad
    Z=\sum_j z_j.
    \label{eq:eigenmode_phasors}
\end{equation}
At zero drive, the QSL steady state is diagonal in the Fock basis. All $n_j$ are integers, all phasors lie on the positive real axis, and $Z=1$. In the weak-drive regime, the eigenvectors acquire only small fractional number shifts and the phasors rotate perturbatively, producing the quadratic GP in \eqref{eq:perturb_GP_harmonic}. Near the critical point, the adjacent population gaps become small, and the drive can reorganize the eigenvectors nonperturbatively. Some fractional parts of $n_j$ then become of order one, so the dominant phasors rotate through finite angles and partially cancel. The sharp drop and the multiple lobes of the GP near the threshold therefore reflect the reorganization and interference of the steady-state eigensystem, rather than a simple change of the mean photon number.

\begin{figure}
    \centering
    \includegraphics[width=0.94\linewidth]{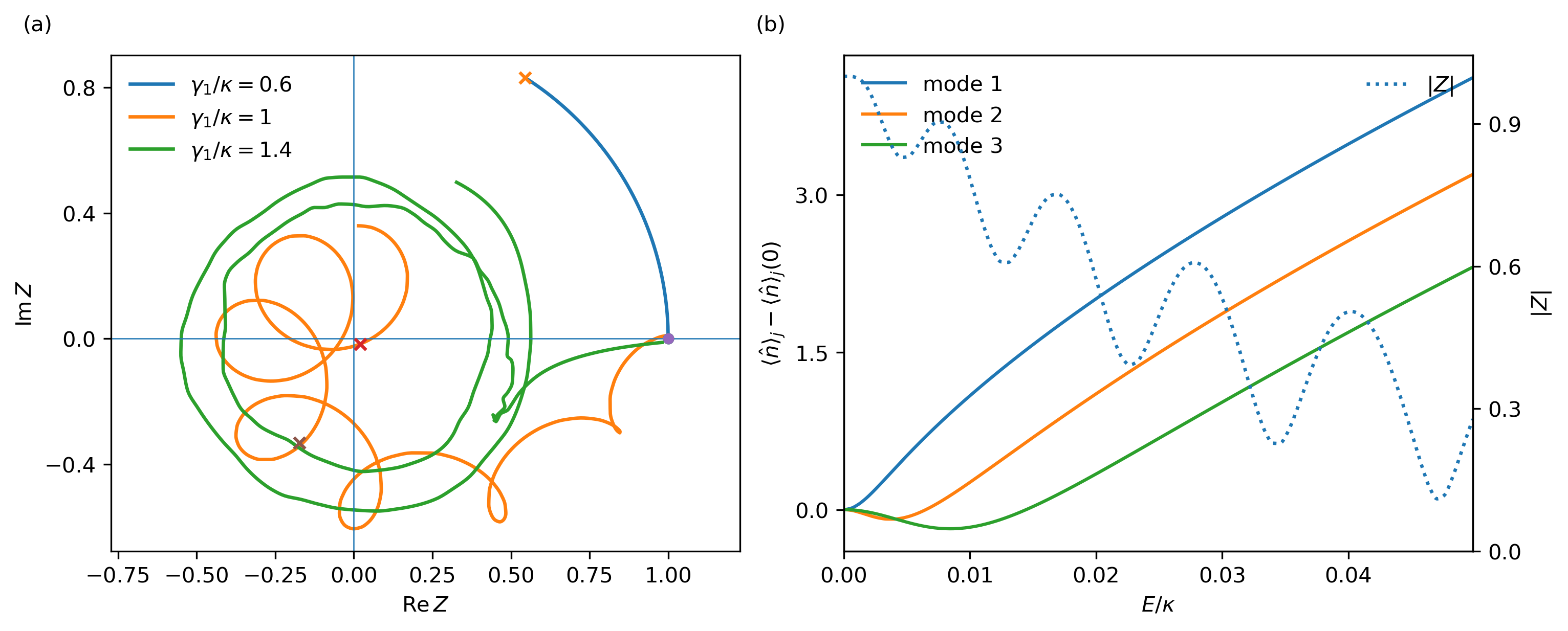}
    \caption{Eigensystem reorganization near the Hopf threshold. Parameters are $\gamma_2/\kappa=0.005$, $\Delta=\eta=U=0$, and $N_{\rm cut}=120$. (a) Complex trajectories of the geometric amplitude for $\gamma_1/\kappa=0.6,1,1.4$ as the harmonic drive increases. Solid dots mark $E=0$ and crosses mark the minimum of $|Z|$ on each trajectory. (b) Fractional number expectations of the dominant tracked eigenmodes on the critical cut $\gamma_1/\kappa=1$}
    \label{fig:hopf_eigensystem}
\end{figure}

\cref{fig:hopf_eigensystem}(a) reveals that $Z$ remains close to the unit circle below the threshold. At the critical point, its trajectory approaches the origin as shown by the cross of the yellow curve, so partial cancellation among the eigenmode contributions strongly enhances the variation of GP. The dominant eigenmode weights remain comparatively smooth, whereas their number expectations develop large drive-induced shifts. The critical response is therefore associated mainly with a nonperturbative deformation of the steady-state eigenvectors and the resulting interference of their geometric contributions, shown by the three solid curves in \cref{fig:hopf_eigensystem}(b). The dotted curve explains the descending oscillation in Fig. 3 (d) and (g) of the main text.

We next consider the two-photon-driven Kerr
resonator~\cite{Bartolo2016,Minganti2018,Beaulieu2025}. In the bright
phase, the steady-state phase-space distribution develops two lobes
centered approximately at $\alpha$ and $-\alpha$, which are related by
parity. An ideal parity-symmetric bright state can therefore be written
as
\begin{equation}
    \rho_{\rm B}
    \simeq
    \frac{1}{2}
    \left(
        |\alpha\rangle\langle\alpha|
        +
        |-\alpha\rangle\langle-\alpha|
    \right)
    =
    w_+|C_+\rangle\langle C_+|
    +
    w_-|C_-\rangle\langle C_-|,
    \label{eq:rho_bright_SM}
\end{equation}
where
\begin{equation}
    |C_\pm\rangle
    =
    \frac{|\alpha\rangle\pm|-\alpha\rangle}
    {\sqrt{2(1\pm e^{-2N_{\rm B}})}},
    \qquad
    N_{\rm B}=|\alpha|^2,
\end{equation}
and
\begin{equation}
    w_\pm
    =
    \frac{1\pm e^{-2N_{\rm B}}}{2}.
\end{equation}
Here $|C_+\rangle$ and $|C_-\rangle$ are the even- and odd-parity cat
states, while $w_+$ and $w_-$ are the corresponding eigenvalues of
$\rho_{\rm B}$.

The number expectations of the two eigenvectors are
\begin{equation}
    n_+
    =
    N_{\rm B}\tanh N_{\rm B},
    \qquad
    n_-
    =
    N_{\rm B}\coth N_{\rm B}.
\end{equation}
The geometric amplitude associated with the ideal bright state is
therefore
\begin{equation}
    Z_{\rm B}
    \simeq
    w_+e^{i2\pi n_+}
    +
    w_-e^{i2\pi n_-}.
    \label{eq:Z_B_SM}
\end{equation}

Near the first-order coexistence region, we introduce a minimal
vacuum--bright decomposition,
\begin{equation}
    \rho_{\rm ss}
    \simeq
    p_{\rm B}^{\rm eff}\rho_{\rm B}
    +
    (1-p_{\rm B}^{\rm eff})\rho_{\rm V},
    \label{eq:rho_two_sector_SM}
\end{equation}
where $\rho_{\rm V}$ denotes the low-occupation, vacuum-like branch.
The corresponding geometric amplitude is
\begin{equation}
    Z_{\rm model}
    \simeq
    p_{\rm B}^{\rm eff}Z_{\rm B}
    +
    (1-p_{\rm B}^{\rm eff})Z_{\rm V}.
    \label{eq:Z_model_SM}
\end{equation}
In the ideal vacuum limit, $Z_{\rm V}\rightarrow1$.

The coefficient $p_{\rm B}^{\rm eff}$ is obtained by matching the first
two normal-ordered moments. For the ideal state in
Eq.~\eqref{eq:rho_two_sector_SM}, with
$\rho_{\rm V}=|0\rangle\langle0|$ and coherent bright components, one
has
\begin{equation}
    \langle\hat n\rangle
    =
    p_{\rm B}^{\rm eff}N_{\rm B},
    \qquad
    \langle a^{\dagger2}a^2\rangle
    =
    p_{\rm B}^{\rm eff}N_{\rm B}^2.
\end{equation}
We therefore define
\begin{equation}
    N_{\rm B}
    =
    \frac{\langle a^{\dagger2}a^2\rangle}
    {\langle\hat n\rangle},
    \qquad
    p_{\rm B}^{\rm eff}
    =
    \frac{\langle\hat n\rangle^2}
    {\langle a^{\dagger2}a^2\rangle}
    =
    \frac{1}{g^{(2)}(0)},
    \label{eq:moment_sector_estimate}
\end{equation}
where
\begin{equation}
    g^{(2)}(0)
    =
    \frac{\langle a^{\dagger2}a^2\rangle}
    {\langle\hat n\rangle^2}.
\end{equation}
The quantity $p_{\rm B}^{\rm eff}$ is thus a moment-matched
bright-sector coefficient. It coincides with the coefficient in \eqref{eq:rho_two_sector_SM} when the steady state is accurately
described by an ideal vacuum component and an incoherent pair of
coherent bright states. In the driven Kerr resonator, it is used as an
effective indicator of the transfer between the bright and
low-occupation branches. 

\begin{figure}
    \centering
    \includegraphics[width=0.96\linewidth]
    {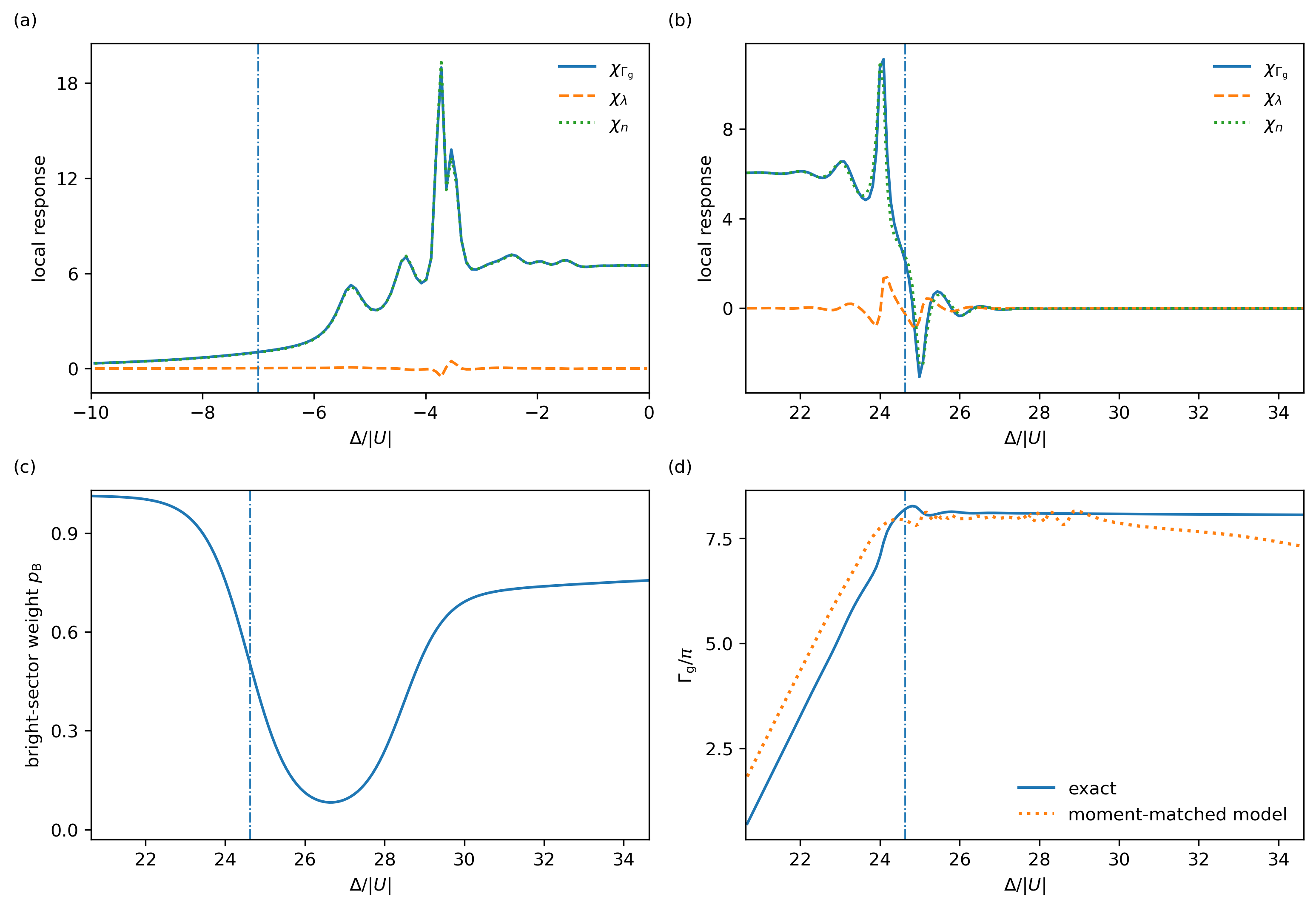}
    \caption{
    Geometric response of the two-photon-driven Kerr resonator across
    the second- and first-order dissipative transition regions.
    Parameters are $U=-1$, $\eta/|U|=3.5$,
    $\gamma_1/|U|=\gamma_2/|U|=\kappa/|U|=0.1$, $E=0$, and
    $N_{\rm cut}=100$.
    (a) Local geometric response near the second-order boundary,
    decomposed as
    $\chi_{\Gamma_{\rm g}}=\chi_\lambda+\chi_n$.
    The dash-dotted vertical line marks the linear instability of the
    vacuum-like branch at $\Delta/|U|=-7$.
    (b) The same decomposition near the first-order boundary. The
    dash-dotted vertical line marks the steepest decrease of the
    steady-state photon number.
    (c) Moment-matched bright-sector coefficient
    $p_{\rm B}^{\rm eff}$ obtained from
    Eq.~\eqref{eq:moment_sector_estimate}.
    (d) Exact unwrapped GP and the vacuum--bright-sector model in the
    first-order region. The model is used only to interpret the
    transfer between the bright and low-occupation branches and does
    not determine either transition boundary.
    }
    \label{fig:kerr_boundaries}
\end{figure}

At the lower boundary, the two $\mathbb{Z}_2$-related bright solutions emerge continuously from the
vacuum-like branch. In the ideal limiting picture, $N_{\rm B}\rightarrow0$, so that
\begin{equation}
    w_+\rightarrow1,
    \qquad
    w_-\rightarrow0,
    \qquad
    n_+\rightarrow0,
    \qquad
    n_-\rightarrow1.
\end{equation}
Consequently,
\begin{equation}
    Z_{\rm B}\rightarrow Z_{\rm V},
\end{equation}
and the GP remains continuous across the transition. The response is characterized directly through
the density-matrix eigensystem. ~\Cref{fig:kerr_boundaries}(a) shows that $\chi_n$ gives the dominant contribution, while $\chi_\lambda$ remains small. The transition therefore primarily
changes the number-space structure of the density-matrix eigenvectors, rather than producing a rapid redistribution of its eigenvalues.

The strongest negative-detuning GP response occurs further inside the bright phase rather than exactly at the second-order boundary. In this region, the eigenmode contributions $\lambda_j e^{i2\pi n_j}$ partially cancel, causing $|Z|$ to become small. The factor $|Z|^{-2}$ in \eqref{eq:chi_gp_SM} then amplifies
the local geometric response. The resulting peak is therefore a
response of the reorganized bright state and should not be identified
with the position of the second-order transition itself.

At the upper boundary, the bright branch retains a finite characteristic occupation while the steady state rapidly switches toward the low-occupation branch. As shown in \cref{fig:kerr_boundaries}(c), $N_{\rm B}$ remains finite while $p_{\rm B}^{\rm eff}$ decreases sharply, which is consistent with a
redistribution of the dominant density-matrix eigenvalues between bright-like and low-occupation eigenmodes rather than a continuous collapse of the bright-state amplitude. This redistribution produces an appreciable contribution $\chi_\lambda$ to the geometric response.
However, \eqref{eq:GP_exact} shows that the phase of $Z$ depends not only on the eigenvalues $\lambda_j$, but also on the number expectations $n_j$ of the corresponding eigenvectors. Consequently, as shown in \cref{fig:kerr_boundaries}(b), $\chi_n$ gives the larger
contribution to the strongest GP excursion, while $\chi_\lambda$ and $\chi_n$ partially cancel near the maximum photon-number slope. The first-order transition is therefore associated with a rapid
redistribution of the density-matrix eigenvalues, but this does not require $\chi_\lambda$ to dominate $\chi_{\Gamma_{\rm g}}$.

The moment-matched model in \cref{fig:kerr_boundaries}(d) provides a qualitative description of the transfer between the bright and low-occupation branches. The actual low-occupation steady state is not an exact vacuum, and the Kerr-deformed bright lobes are not exact coherent states. Moreover, the extracted $N_{\rm B}$ enters $Z_{\rm model}$ through the rapidly varying factors
$e^{i2\pi n_\pm}$, so a small difference in $N_{\rm B}$ can produce a large difference in the predicted GP. We therefore use \eqref{eq:rho_two_sector_SM}--\eqref{eq:moment_sector_estimate} only to interpret the transition. All quantitative values of the GP, $\chi_\lambda$, and $\chi_n$ are calculated from the complete steady-state density matrix and its eigenvalues and eigenvectors.

\section{Platform-tailored interferometric determination of the GP}
\label{sec:experimental_gp}

The covariance relation in \eqref{eq:ss_rotation} implies that the
steady-state trajectory is a unitary orbit, even though
$\rho_{\rm ss}(0)$ is prepared dissipatively. The GP can therefore be
determined by mixed-state interferometry without reconstructing the
complete density matrix at every parameter point
\cite{Singh_2003, Daniel2023}.

For a nondegenerate steady state, a parallel-transport representative
of the orbit is
\begin{equation}
    U_{\rm PT}(\phi)
    =
    R(\phi)F(\phi),
    \qquad
    F(\phi)
    =
    \sum_j
    e^{i\phi\langle\hat n\rangle_j}
    |\psi_j\rangle\langle\psi_j|.
    \label{eq:U_PT}
\end{equation}
Since $F(\phi)$ commutes with $\rho_{\rm ss}(0)$,
$U_{\rm PT}(\phi)$ generates the same density-matrix trajectory as
\eqref{eq:ss_rotation}, while each eigenvector satisfies
\begin{equation}
    \langle\psi_j|
    U_{\rm PT}^\dagger(\phi)
    \partial_\phi U_{\rm PT}(\phi)
    |\psi_j\rangle
    =
    0.
    \label{eq:PT_condition}
\end{equation}
At the end of the loop,
\begin{equation}
    \operatorname{Tr}
    \left[
        \rho_{\rm ss}(0)U_{\rm PT}(2\pi)
    \right]
    =
    Z,
    \label{eq:interferometric_Z}
\end{equation}
with $Z$ defined in \eqref{eq:GP_exact}. A bare number rotation would
not suffice because $R(2\pi)=\mathbbm{1}$.

An ancilla-assisted Ramsey sequence can measure
\eqref{eq:interferometric_Z}. Preparing the ancilla in $|+\rangle$ and
applying the controlled operation
\begin{equation}
    C[U_{\rm PT}]
    =
    |0\rangle_{\rm a}\langle0|\otimes\mathbbm{1}
    +
    |1\rangle_{\rm a}\langle1|\otimes U_{\rm PT}(2\pi)
    \label{eq:controlled_U_PT}
\end{equation}
gives the Ramsey signal
\begin{equation}
    P_{\rm a}(\chi)
    =
    \frac{1}{2}
    \left[
        1+
        \operatorname{Re}
        \left(
            e^{-i\chi}Z
        \right)
    \right]
    =
    \frac{1}{2}
    \left[
        1+
        |Z|\cos(\chi-\Gamma_{\rm g})
    \right].
    \label{eq:Ramsey_signal}
\end{equation}
Measurements at $\chi=0$ and $\chi=\pi/2$ determine
$\operatorname{Re}Z$ and $\operatorname{Im}Z$, respectively. Thus,
the phase shift and fringe visibility yield
$\Gamma_{\rm g}=\arg Z$ and $\nu_{\rm g}=|Z|$.

\subsection{Trapped-ion QSL oscillator}

For the trapped-ion realization of the QSL oscillator
\cite{Li2025}, the motional mode carries $\rho_{\rm ss}$ and an
internal electronic transition acts as the Ramsey ancilla. The
steady state is first prepared using the engineered gain, nonlinear
damping, and synchronization drive. These dissipative processes are
then switched off during the coherent interferometric sequence.

In the deep quantum regime, the relevant steady-state eigensystem is
well approximated by one two-level block and a spectator Fock state.
For the block spanned by $|0\rangle$ and $|q\rangle$, with $q=1$ for
harmonic driving and $q=2$ for squeezing, write
\begin{equation}
    \rho_{0q}
    =
    \frac{\Sigma_{0q}}{2}\mathbbm{1}
    +
    \frac{1}{2}
    \bm r_{0q}\cdot\bm\sigma,
    \qquad
    \bm r_{0q}
    =
    \left(
        2\operatorname{Re}\rho_{0q},
        -2\operatorname{Im}\rho_{0q},
        d_{0q}
    \right).
    \label{eq:ion_block}
\end{equation}
Let $B_{0q}$ diagonalize this block,
\begin{equation}
    B_{0q}
    \left(
        \hat{\bm r}_{0q}\cdot\bm\sigma
    \right)
    B_{0q}^\dagger
    =
    \sigma_z.
    \label{eq:ion_basis_rotation}
\end{equation}
Within the three-level approximation, the required final operation is
then
\begin{equation}
    U_{\rm PT}^{(0q)}(2\pi)
    =
    B_{0q}^\dagger
    \begin{pmatrix}
        e^{i2\pi n_+^{(0q)}} & 0\\
        0 & e^{i2\pi n_-^{(0q)}}
    \end{pmatrix}
    B_{0q}
    \oplus\mathbbm{1}_{\perp},
    \label{eq:ion_U_PT}
\end{equation}
where the number expectations are given by
\eqref{eq:n_three_level_harmonic} or
\eqref{eq:n_three_level_squeezing}.

The controlled number rotation may be generated by an effective
state-dependent motional interaction,
\begin{equation}
    H_{\rm ion}
    =
    \hbar\chi_{\rm ion}
    |e\rangle\langle e|
    \otimes\hat n,
    \label{eq:ion_control}
\end{equation}
while carrier and sideband pulses provide the basis rotations
$B_{0q}$. The GP Arnold tongue and the Hopf-critical response can then
be obtained by repeating the two Ramsey measurements in
\eqref{eq:Ramsey_signal} as the detuning, drive strength, or effective
linear damping is varied.

Only the small block in \eqref{eq:ion_block} must be calibrated.
Full motional tomography is therefore replaced by a few-level
calibration followed by two ancilla measurements per parameter point.

\subsection{Two-photon-driven Kerr resonator}

For the Kerr-resonator DPT protocol, the nonlinear resonator carries
$\rho_{\rm ss}$ and a dispersively coupled transmon provides the
ancilla. The two-photon drive is first applied until the resonator
reaches its steady state at the selected $(\Delta,\eta)$. The
stabilizing dynamics are then interrupted during the Ramsey sequence.

The dispersive interaction
\begin{equation}
    H_{\rm disp}
    =
    \frac{\hbar\chi_{\rm c}}{2}
    \hat n\sigma_z
    \label{eq:dispersive_control}
\end{equation}
provides the controlled number rotation. Inside the bright phase, the
dominant density-matrix eigenvectors are approximately the even- and
odd-parity cat-like modes introduced in
\eqref{eq:rho_bright_SM}. In the reduced vacuum--cat manifold, the
parallel-transport operation takes the approximate form
\begin{equation}
    U_{\rm PT}^{\rm Kerr}(2\pi)
    \simeq
    e^{i2\pi n_{\rm V}}
    |\psi_{\rm V}\rangle\langle\psi_{\rm V}|
    +
    e^{i2\pi n_+}
    |C_+\rangle\langle C_+|
    +
    e^{i2\pi n_-}
    |C_-\rangle\langle C_-|,
    \label{eq:kerr_U_PT}
\end{equation}
where $n_\pm$ are given below \eqref{eq:rho_bright_SM} and
$n_{\rm V}\simeq0$ for the vacuum-like mode.

Parity mapping and number-selective control can be used to transform
the three modes in \eqref{eq:kerr_U_PT} to a controllable basis,
apply the ancilla-conditioned phases, and reverse the transformation.
The Ramsey signal in \eqref{eq:Ramsey_signal} then gives the GP across
both DPT boundaries. The accompanying visibility $|Z|$ also measures
the cancellation among the dominant geometric phasors and is expected
to decrease inside regions of strong eigenmode interference.

Near the first-order coexistence region, the exact eigenmodes need not
be ideal vacuum and cat states. Partial Wigner or parity measurements
can therefore be used to calibrate the reduced manifold at a few
representative points. Full tomography is retained as a benchmark,
rather than being required throughout the complete detuning scan.

The interferometric protocol replaces the measurement cost of full
tomography by the coherent-control cost of implementing
$U_{\rm PT}$. It is therefore most favorable in the few-level
trapped-ion regime and in the reduced vacuum--cat manifold of the
Kerr resonator. If exact density-matrix degeneracies occur, the
Abelian construction in \eqref{eq:U_PT} must be replaced by the
corresponding block parallel transport within each degenerate
eigenspace \cite{Singh_2003}.

\bibliographystyle{unsrt} 
\bibliography{supple_ref}